\documentclass{aastex} 
\usepackage{emulateapj5, apjfonts, psfig, epsfig} 
\begin{document}

\slugcomment{ApJL, in press}

\title{Anomalous Evolution of the Dwarf Galaxy 
H{\sc i}PASS~J1321-31\footnotemark}
\footnotetext{Based on observations with the NASA/ESA {\it Hubble Space 
Telescope}, obtained at the Space Telescope Science Institute, which 
is operated by the Association of Universities for Research in Astronomy, 
Inc., (AURA), under NASA Contract NAS 5-26555.}

\shortauthors{Pritzl et al.}
\shorttitle{H{\sc i}PASS~J1321-31 dwarf galaxy}

\received{}
\accepted{}
\revised{}

\author{Barton J. Pritzl\altaffilmark{2}, Patricia M. Knezek\altaffilmark{3}, 
John S. Gallagher, III\altaffilmark{4}, Marco Grossi\altaffilmark{5}, 
Mike J. Disney\altaffilmark{5}, Robert F. Minchin\altaffilmark{5}, 
Kenneth C. Freeman\altaffilmark{6}, Eline Tolstoy\altaffilmark{7}, 
and Abi Saha\altaffilmark{2}} 

\altaffiltext{2}{National Optical Astronomy Observatory, P.O. Box 26732, 
Tucson, AZ 85726; pritzl@noao.edu, asaha@noao.edu} 
\altaffiltext{3}{WIYN, Consortium, Inc., P.O. Box 26732, Tucson, AZ 85726; knezek@wiyn.org} 
\altaffiltext{4}{Department of Astronomy, University of Wisconsin, Madison, WI 
53706-1582; jsg@astro.wisc.edu} 
\altaffiltext{5}{University of Wales, Cardiff, P.O. Box 913, Cardiff, U.K., 
CF2 3YB; Marco.Grossi@astro.cf.ac.uk, mjd@astro.cf.ac.uk, 
Robert.Minchin@astro.cf.ac.uk} 
\altaffiltext{6}{Research School of Astronomy and Astrophysics, Mount Stromlo 
Observatory, Cotter Road, Weston, ACT 2611, Australia; kcf@mso.anu.edu.au} 
\altaffiltext{7}{Kapteyn Institute, University of Groningen; 
etolstoy@astro.rug.nl}

\begin{abstract} 

We present HST/WFPC2 observations of the dwarf galaxy H{\sc i}PASS~J1321-31.  
This unusual galaxy lies in the direction of the Centaurus~A group of 
galaxies, and has a color-magnitude diagram 
with a distinctive red plume of luminous stars.  This 
feature could arise from (a) a red giant branch if the galaxy were much 
nearer than previously recognized, (b) a peculiar asymptotic giant branch, 
or, (c) an $\sim 1$ Gigayear old population of intermediate mass 
red supergiants, which we find to be the most likely explanation.  However, 
the lack of equally luminous blue stars requires that the star formation 
has dropped substantially since these stars were formed.  Evidently 
H{\sc i}PASS~J1321-31 experienced an episode of enhanced star formation 
rather recently in its star formation history followed by a period of 
relative quiescence which has led to the evolution of the main sequence 
stars into the red supergiant branch.  The stellar populations in 
H{\sc i}PASS~J1321-31 reflect a star formation history that is uncommon 
in star forming dwarf galaxies.  This is the first time such a star 
formation history has been noted, although the literature contains a 
small number of other dwarf galaxies with similar color-magnitude 
diagrams.  Therefore, H{\sc i}PASS~J1321-31 and these other galaxies 
represent a different path of dwarf galaxy evolution that has not been 
well-explored and an important probe into how dwarf galaxies evolve.

\end{abstract}


\section{Introduction} 

Dwarf galaxies are the most common type of galaxy in the Universe. 
In cosmological models, such as those invoking cold dark matter, 
the first and most frequently formed gravitationally bound objects 
resemble present-day dwarfs.  Many studies have been made of these galaxies, 
allowing us a glimpse into the conditions and evolution of galaxies 
in the youthful Universe.  While the stellar populations of dwarf galaxies 
in the Local Group have been extensively investigated (see Mateo 1998 and 
references within; Grebel 1997), this represents only one environment in 
which these types of galaxies are known to exist.  Still
open is the question of how, and if, dwarf galaxies evolve from initial gas 
clouds to dwarf irregular galaxies to dwarf spheroidal galaxies (e.g., 
Grebel, Gallagher, \& Harbeck 2003).  Furthermore, we need to establish 
whether dwarf galaxies outside the Local Group have similar relationships 
between star formation histories and other global and environmental 
properties. 

\begin{figure*}[t]
  \centerline{\psfig{figure=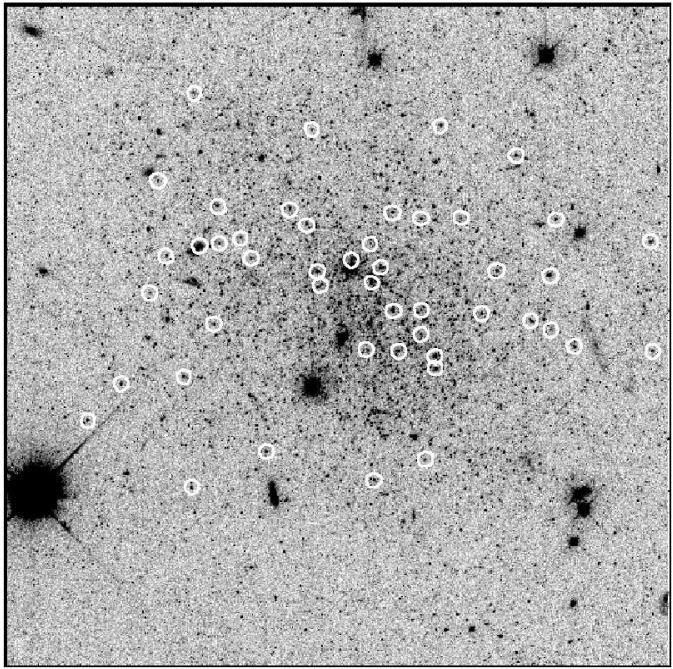,height=3.50in,width=3.50in}}
  \caption{Image of H{\sc i}PASS~J1321-31 from the WFC3 
           showing its well-resolved stellar population.  The circles 
           indicate the stars that belong to the red plume seen in Figure~2a.} 
  \label{Fig01} 
\end{figure*}

Numerous studies have explored the global properties of dwarf galaxies in 
nearby galaxy groups and clusters.  However, detailed investigations of star 
formation histories derived from resolved stellar populations are 
limited to the nearest galaxy groups, which require the 
capabilities of the {\it Hubble Space Telescope} (HST).  Thus we now have 
Wide-Field Planetary Camera~2 (WFPC2) color-magnitude diagrams (CMDs) for 
significant samples of dwarf galaxies in the M81 (e.g., Karachentsev et al.\ 
2002a) and the Centaurus~A (Cen~A; Karachentsev et al.\ 2002b) groups.  
These and similar data suggest diverse star formation histories among 
nearby dwarfs, similar to those found in the Local Group.  Dwarf spheroidal 
and transition galaxies show dominant red giant branches (RGBs), while the 
luminous stellar populations in irregulars consist of a mixture of the RGB 
stars along with more massive stars and their evolved descendants.  Galaxies 
where the old RGB is weak are rare, but this condition is seen in the 
dwarf irregular galaxies Leo~A (Tolstoy et al.\ 1998; Schulte-Ladbeck 
et al.\ 2002) and Holmberg~IX (Karachentsev et al.\ 2002a; Marakova et al.\ 
2002).  This suggests that larger samples of resolved dwarf 
galaxies will reveal rarer or shorter lived evolutionary 
paths. 

One property that makes the dwarf galaxies in Cen~A interesting to study 
is the fact that so many of them are gas-rich.  No dwarfs in the Local 
Group are young in the sense that they still contain more mass in 
gas than in stars.  However, some of those that have been found in a blind 
H{\sc i} survey of the nearby Cen~A group are (H{\sc i} Parkes 
All-Sky Survey [H{\sc i}PASS]; Banks et al.\ 1999).  After a number of new 
candidate dwarf galaxies were discovered, we undertook follow-up observations 
of a select number of them using the HST/WFPC2, giving us the opportunity to 
study the resolved stellar populations and thus the evolutionary histories 
in gas-rich Cen~A group dwarfs.  In this letter we discuss 
the unusual nature of one of these systems, H{\sc i}PASS~J1321-31.  In a 
future paper we discuss its physical properties along with the other 
Cen~A dwarfs observed is this program, H{\sc i}PASS~J1337-39 and 
H{\sc i}DEEP~J1337-33 (Grossi et al.\ 2003).  We find that the 
color-magnitude diagram (CMD) of H{\sc i}PASS~J1321-31 shows a significantly 
different stellar population from the other two dwarf galaxies 
surveyed, as well as most Local Group dwarfs.  In the following we present 
the CMD of H{\sc i}PASS~J1321-31 and discuss the possible origins of its 
``unique'' stellar populations.

\begin{figure*}[t]
  \centerline{\psfig{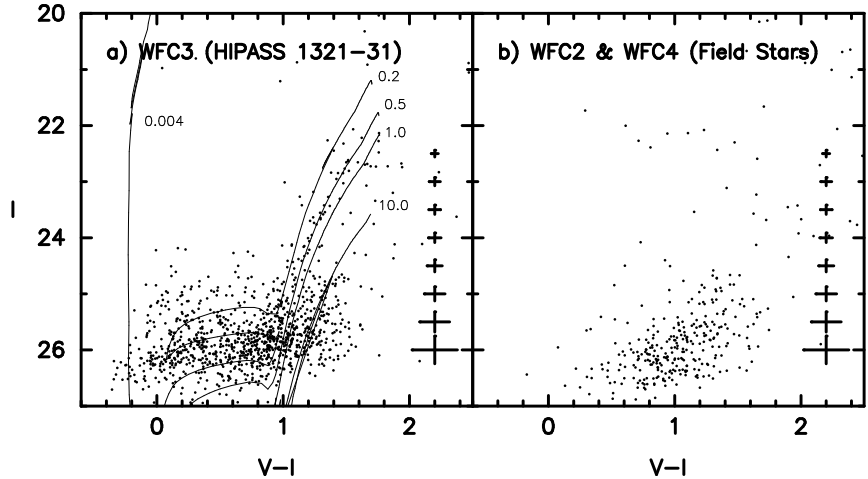}}
  \caption{Color-magnitude diagrams for (a) the WFC3 which 
           shows H{\sc i}PASS~J1321-31 and (b) the field surrounding the 
           galaxy from the WFC2 and WFC4.  A comparison of the two plots 
           shows that H{\sc i}PASS~J1321-31 contains a population of blue 
           stars and a red plume of stars.  Padova theoretical isochrones 
           (Girardi et al.\ 2002) are shown for $Z=0.0004$, where 
           the ages listed are in Gyr.  In this case, the best fit to the 
           red supergiant branch is 0.5~Gyr.  Isochrones showing the young 
           (0.004~Gyr) and old (10~Gyr) tracks are for illustrative purposes.} 
  \label{Fig02}
\end{figure*}

\section{Observations and Reductions} 

H{\sc i}PASS~J1321-31 was observed with the WFPC2 on 2001 June 12 with 
four exposures in F555W at 1200~sec each and four exposures in F814W 
at 1300~sec each, with the galaxy centered on the WFC3 (Figure~1).  The 
images for each filter and chip were combined to create a deeper image 
of the galaxy.  {\sc daophot}, {\sc allstar}, and {\sc allframe} were 
used to obtain the profile-fitted photometry of the stars presented in this 
letter.  The point spread functions (PSFs) used in the reductions were kindly 
communicated to us by P. B. Stetson (see Stetson et al.\ 1998).  Due to the 
lack of stars on the PC, we chose not to reduce that chip.  Limits in $\chi$
($\chi > 2.0$) and sharpness ($-0.1 < {\rm sharp} < 0.1$) were 
placed on the fits of the stars to remove any objects with questionable 
photometry.  CTE corrections were made using the equations from 
Whitmore, Heyer, \& Casertano (1999).  After the aperture corrections were 
calculated from the unsaturated brightest stars in each chip, the photometry 
was calibrated to the standard $V$ and $I$ system using the 
equations in Holtzman et al.\ (1995). 

To check the reliability of the photometry and confirm that the red plume 
apparent in the CMD was ``real," the data were also reduced using DoPHOT and 
independently using {\sc daophot/allstar} with different PSFs.  In both cases, 
the red plume was still present.  We also checked the location of the red 
plume stars against any defects in the WFC3, such as warm pixels, and found 
the photometry was unaffected by these difficulties.

\section{Color-Magnitude Diagram} 

Figure~2a shows the CMD for the WFC3, which was centered on 
H{\sc i}PASS~J1321-31.  The CMD for the field (WFC2 and WFC4) is seen in 
Figure~2b.  Figure~2a shows a clear red plume of luminous stars.  These 
stars are scattered across the optical extent of the galaxy (see Figure~1).  
We know that H{\sc i}PASS~J1321-31 has a heliocentric 
velocity of 572~km/s according to the H{\sc i} survey of Banks et al.\ (1999).  
This is consistent with it being a member of the Cen~A Group and proves to 
be important in interpreting the CMD.\@  In the following we discuss three 
possible interpretations for the red plume:  A RGB, an asymptotic giant 
branch (AGB), or a red supergiant branch (RSGB).

\subsection{The Red Plume} 

One might think that the red plume seen in Figure~2a is 
the RGB of H{\sc i}PASS~J1321-31.  Assuming this to be true, 
the tip of the red giant branch (TRGB) would be at $m_I\sim22.6$~mag.  
In this case using this value for the TRGB, we estimate the distance to 
the galaxy to be about 2.0~Mpc, assuming $M_{I,{\rm TRGB}}=-4.05$ 
(Da~Costa \& Armandroff 1990) and $E(\bv)=0.062\pm0.01$ (Schlegel, 
Finkbeiner, \& Davis 1998).  This estimate places H{\sc i}PASS~J1321-31 
just outside the Local Group in the direction of the Cen~A group of galaxies.  
The distance to the giant elliptical galaxy Cen~A is $3.9\pm0.3$~Mpc (Harris, 
Harris, \& Poole 1999), but H{\sc i}PASS~J1321-31 is located closer in the sky 
to M83, which has a distance of $4.5\pm0.3$~Mpc (Thim et al.\ 2003).  This 
interpretation would place H{\sc i}PASS~J1321-31 somewhere between the Milky 
Way and M83.  This is in conflict with the radial velocity found for the 
H{\sc i} cloud associated with the galaxy, which is consistent with it 
being a member of the more distant Cen~A group.  This apparent conflict 
could be resolved if the H{\sc i} cloud were not associated with the 
stellar galaxy, but such a chance coincidence would be unique and is 
highly unlikely. 

However, the biggest challenge to the RGB interpretation is the fact that 
the red plume is not well populated even well below its tip.  For globular 
clusters and dwarf galaxies, the number of stars increases as one moves 
fainter along the RGB.\@  Figure~3 shows the luminosity function for 
H{\sc i}PASS~J1321-31.  The luminosity function has been Gaussian-smoothed 
using Eq.~A1 in Sakai, Madore, \& Freedman (1996).  In the magnitude range 
from $24.0 < m_I < 22.0$, the luminosity function is relatively flat.  
We used the data in Saviane, Held \& Piotto (1996) to create a luminosity 
function for the RGB in Tucana (the dashed line in Figure~3).  The tip of 
the Tucana RGB was shifted to match the tip of the red plume in 
H{\sc i}PASS~J1321-31 and was scaled so that the upper luminosity functions 
are reasonably well-matched.  The relatively constant luminosity function 
of the H{\sc i}PASS~J1321-31 red plume is in contrast to the increasing 
luminosity function of the Tucana RGB.  The combination of distance 
difficulties and flat luminosity function argue that the red plume is 
not a RGB.\@ 

A second possibility is that the red plume is an AGB.\@  A strong AGB 
population is seen in dwarf galaxies like IC~1613 (Cole et al.\ 1999) 
and Sextans~A (Dohm-Palmer et al.\ 2002).  In these galaxies, the 
AGB extends around 0.5~mag in $I$ above the TRGB and then turns redward.  
This differs from the behavior of the red plume in 
H{\sc i}PASS~J1321-31. It extends at least two 
magnitudes above any possible TRGB and begins its assent from the 
bluer part of the possible RGB.\@  Thus, it is implausible that the 
red plume consists of AGB stars. 

Ruling out the RGB and AGB leaves us with the likelihood that these stars 
make up the RSGB.\@  Such stars originate from intermediate mass 
stars in post-main sequence helium core burning evolutionary phase, the 
same phases that produce pronounced ``blue loops" at low metallicities.  
A beautiful example of this phenomenon is seen in Sextans~A, where stars with 
ages of $<1$~Gyr mark the red and blue edges of loop evolution 
(Dohm-Palmer et al.\ 1997).  However, for cases where the evolved stars 
have ages near 1~Gyr the red branch of the evolutionary loop dominates 
(see Fig.~3 of Dohm-Palmer et al.).  Thus in this model H{\sc i}PASS~J1321-31 
would be in something resembling a post burst phase where the star formation 
rate was enhanced $\le 1$~Gyr in the past, giving rise to an unusual number 
of evolved stars with initial masses slightly less than $3~M_{\odot}$. 

We searched the literature for other dwarf galaxies with similar CMDs 
as H{\sc i}PASS~J1321-31.  In the HST snapshot survey of dwarf galaxies 
(Karachentsev et al.\ 2002a, b, c; 2003a, b, c) there are several with red 
plumes similar to H{\sc i}PASS~J1321-31.  They are all found to 
extend from the bluer part of the RGB and trend slightly to the red as one 
moves to higher luminosity along the branch.  We investigated the number of 
stars along the red plumes in a couple of these galaxies and found that 
they remain relatively constant, similar to what we 
see in the luminosity function for H{\sc i}PASS~J1321-31 (see Figure~3).  
Therefore, as noted by Karachentsev et al., it is likely these red plumes 
are RSGBs, including the one in H{\sc i}PASS~J1321-31, arising from 
intermediate mass stars. 

\begin{figure*}[t]
  \centerline{\psfig{figure=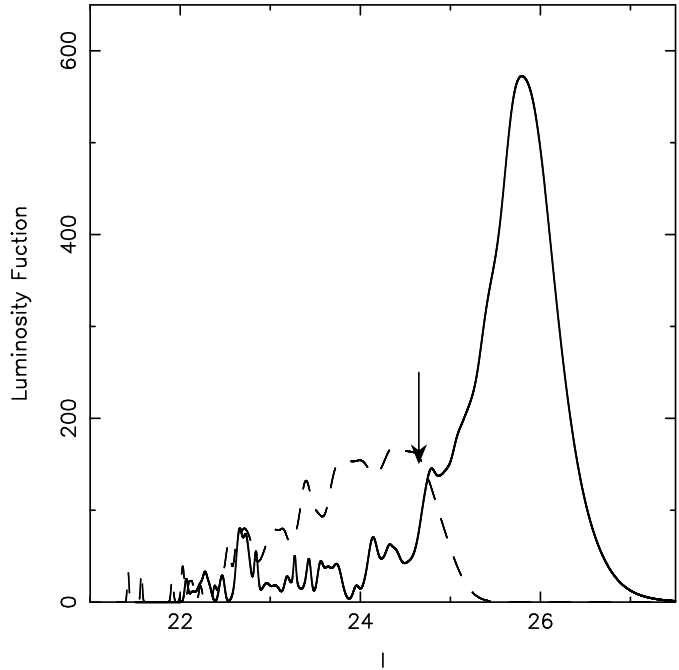,height=3.50in,width=3.50in}}
  \caption{Luminosity function of the H{\sc i}PASS~J1321-31 
           stars within $0.8 < V-I < 2.0$ as shown by the solid line.  
           It has been Gaussian-smoothed using Eq.\ A1 in Sakai, Madore, 
           \& Freedman (1996).  The dashed line represents the luminosity 
           function of the Tucana dwarf galaxy's red giant branch.  The tip 
           of the Tucana red giant branch was shifted to match the tip of 
           the red plume in H{\sc i}PASS~J1321-31.  This illustrates the 
           constant luminosity function of the red plumes, in contrast to 
           the increasing luminosity function of a red giant branch.  The 
           arrow indicates the location of the tip of the red giant 
           branch in H{\sc i}PASS~J1321-31.} 
  \label{Fig03} 
\end{figure*}

We can test the intermediate mass RSG hypothesis by seeing if the presence 
of a normal RGB is consistent with our data.  Following the approach 
adopted in the Karachentsev et al.\ papers, we used the edge-detection 
filter equation in Sakai, Madore, \& Freedman (1996) to search for the 
TRGB.\@  Applying the equation to those stars between $0.8 < V-I < 2.0$ 
results in an edge being detected at $m_I=24.65\pm0.11$~mag, where the 
error includes the photometric uncertainty ($\pm0.06$), the uncertainty 
in the TRGB estimate ($\pm0.06$), the uncertainty of the photometric 
zeropoint ($\pm0.05$), and the uncertainty in the aperture corrections 
($\pm0.05$).  This agrees with the visual impression of the point where 
the luminosity function begins to rise in Figure~3, and is a clear 
signature of the beginning of the RGB.\@  Assuming the TRGB to be at 
$M_I=-4.05$ (Da~Costa \& Armandroff 1990), $E(\bv)=0.062\pm0.01$ 
(Schlegel, Finkbeiner, \& Davis 1998), and $R_V=3.1$, the distance modulus 
estimated from the TRGB is $(m-M)_0=28.59\pm0.13$.  The distance to 
H{\sc i}PASS~J1321-31 is thus $5.2\pm0.3$~Mpc.  This would place the 
galaxy beyond M83, but it would still be a member of the Cen~A group, 
in good agreement with the heliocentric velocity of the H{\sc i} cloud.

\section{What Does This Mean?} 

What are the astrophysical implications of such a pronounced RSGB in 
H{\sc i}PASS~J1321-31 and other dwarfs?  The difficulty in understanding 
the RSGB in H{\sc i}PASS~J1321-31 is the apparent deficiency of corresponding 
BSGB stars.  Dwarf galaxies such as Sextans~A, IC~1613, and those observed 
in the HST snapshot survey for dwarf galaxies, all exhibit a strong 
population of blue stars reaching nearly as high as the RSGB.\@  These 
blue stars are made up of an uncertain combination of upper main sequence 
and evolved intermediate mass stars with ages $<300$~Myr.  The comparable 
luminosity of the red and blue plume in Sextans~A and IC~1613 then 
reflects relatively high levels of ongoing star 
formation.  The absence of bright blue stars in H{\sc i}PASS~J1321-31 cannot 
be due to observational problems, for we see them in the HST data of the 
other H{\sc i}PASS dwarf galaxies (Grossi et al.\ 2003). 

One possible scenario is that H{\sc i}PASS~J1321-31 experienced 
an episode of enhanced star formation, and possibly a mild starburst, 
less than a Gyr ago when the stars now populating the RSGB were formed.  
Using the Bertelli et al.\ (1994) and Girardi et al.\ (2002) isochrones, 
we found that evolved stars with masses around 2-3 $M_{\odot}$ and low 
metallicities on the red branch of the core-helium burning phase would be 
dominant, while the corresponding blue loop stars would be too faint to 
be detected in our observations.  The presence of the faint blue hump 
implies that star formation continued for a few hundred million years at a 
lower rate.  Figure~2a shows a number of theoretical isochrones (Girardi 
et al.\ 2002) for $Z=0.0004$.  While this gives some idea of the age range 
of the RSGB, more detailed modeling will be discussed in Grossi et al.\ 
(2003).  In addition to an absence of luminous blue plume stars, we did 
not find any H{\sc ii} regions in narrow-band H$\alpha$ images taken with 
the WIYN telescope.  H{\sc i}PASS~J1321-31 has a high amount of gas in it 
($M_{HI}/L_B=5.0$; Table~1).  If it did experience a significant burst of 
star formation in its past, this could have spurred further star formation 
throughout the galaxy.  Yet, there is no evidence of recent or continuous 
star formation.  An interesting question then is not only why the star 
formation rose in the past but also why it now appears to have fallen? 

We find the tip of the RSGB to be around $m_I=22.6$~mag yielding an 
absolute magnitude of $M_I=-6.1$~mag assuming $M_I=-4.05$ for the TRGB.\@  
The Karachentsev et al.\ (2002a, b, c; 2003a, b, c) galaxies with 
well-populated normal RSGBs have tips from $M_I=-7.0$ up to $-8.7$.  So 
the faintest RSGB tip in these other galaxies is still about one magnitude 
brighter than that for H{\sc i}PASS~J1321-31.  This suggests that its 
RSGB derives from an older stellar population since the younger stars 
which would populate the brighter end of the RSGB are gone, i.e., evolved 
away, and the corresponding BSGB stars are missing as well.  Therefore, 
H{\sc i}PASS~J1321-31 shows a star formation history that is uncommon in 
known dwarf galaxies. 

Searching through the Karachentsev et al.\ (2002a, b, c; 2003a, b, c) 
papers, we have found only two galaxies with similar CMDs to 
H{\sc i}PASS~J1321-31 whose observed properties we list in Table~1.  None 
of the Local Group dwarfs have this feature in their CMDs.  ESO~444-G084 
(Karachentsev et al.\ 2002b), another galaxy in the Cen~A group, shows a 
possible RSGB which is less populated than the one in H{\sc i}PASS~J1321-31, 
along with a fainter population of blue stars.  Another nearby dwarf 
galaxy KK~65 (Karachentsev et al.\ 2003a) has a CMD with what looks to be 
a dispersed RSGB.  A faint population of blue stars can also be seen, but 
the difference between the tip of the RSGB and the blue plume is not as 
great as that in H{\sc i}PASS~J1321-31.  From Table~1 we can see that 
each of these galaxies are gas-rich, but H{\sc i}PASS~J1321-31 stands 
apart due to its low luminosity.  Another dwarf galaxy, UGC~8833, found 
in the Canes Venatici~I cloud has a similar blue plume and RGB to 
H{\sc i}PASS~J1321-31, but it lacks a RSGB (Karachentsev et 
al.\ 2003b).  The physical properties of these two galaxies are similar, 
except that H{\sc i}PASS~J1321-31 has a much higher H{\sc i} mass-to-light 
ratio (see Table~1).  This comparison helps illustrate that 
H{\sc i}PASS~J1321-31 must have had a period of increased star formation 
in its rather recent past in order to create its RSGB.

\section{Summary \& Conclusions} 

We present the CMD for the gas-rich dwarf galaxy H{\sc i}PASS~J1321-31 as 
observed by the HST/WFPC2.\@  A bright red plume of stars along with a 
much fainter ($\sim 2$~mag) population of blue stars indicate that the 
galaxy has experienced a peculiar star formation history.  Three 
explanations for this red plume were discussed.  We rule out the 
possibilities of it being a RGB or AGB, and conclude that the red plume is 
composed of core helium-burning post-main sequence stars with ages 
{$\lesssim$ 1~Gyr}, possibly as young as 0.4 Gyr.  These are likely to be 
related to an epoch when the galaxy experienced an increase in the star 
formation rate.  This event was not associated with the birth of the 
galaxy since an older RGB is present --- H{\sc i}PASS~J1321-31 is not a 
young galaxy. 
Several puzzles remain, 
such as the source of the possible starburst and how the large gas content 
and unusual evolutionary history might be linked (e.g., could 
this dwarf consists of tidal debris? see Marakova et al.\ 2002).

\acknowledgements 

This research was supported in part by NASA through grant number GO-09115.07 
from the Space Telescope Science Institute, which is operated by AURA, Inc., 
under NASA contract NAS 5-26555.  MG and RFM acknowledge the support of the 
UK Particle Physics and Astronomy Research Council.  Thanks to P. B. Stetson 
for sharing his PSFs for the WFPC2 and his data reduction programs.

\begin{deluxetable}{ccccccccc} 
\tablewidth{0pc} 
\tablecaption{Physical Properties of H{\sc i}PASS~J1321-31 and Comparison 
Dwarf Galaxies \label{tbl-1}} 
\tablehead{
\colhead{Galaxy} & \colhead{RA} & \colhead{Dec} & \colhead{$v_{\odot}$} & 
\colhead{$a$} & \colhead{$m_B$} & \colhead{$M_B$} & \colhead{$M_{\rm HI}$} & 
\colhead{$M_{\rm HI}/L_B$} \\ 
 & \colhead{(2000)} & \colhead{(2000)} & \colhead{km/s} & \colhead {\arcmin} 
& & & \colhead{$10^7 M_{\odot}$} & 
          } 
\startdata 
H{\sc i}PASS J1321-31 & 13:21:06 & --31:32:25 & 572 & 1.2 & 17.1 & --11.7 & 3.7 & 5.0 \\ 
ESO 444-G084          & 13:37:20 & --28:02:45 & 591 & 4.6 & 15.1 & --13.5 & 5.6 & 1.4 \\ 
KK 65                & 07:42:32 &  +16:33:39 & 554 & 4.5 & 15.6 & --12.7 & 1.6 & 0.9 \\ 
UGC 8833             & 13:54:49 &  +35:50:15 & 228 & 0.9 & 15.2 & --12.4 & 1.3 & 1.0 \\
\enddata 
\end{deluxetable} 

\end{document}